\documentclass{article}

\usepackage{PRIMEarxiv}

\usepackage[utf8]{inputenc} 
\usepackage[T1]{fontenc}    
\usepackage{hyperref}       
\usepackage{url}            
\usepackage{booktabs}       
\usepackage{amsfonts}       
\usepackage{nicefrac}       
\usepackage{microtype}      
\usepackage{lipsum}
\usepackage{cleveref}
\usepackage{float}
\usepackage{fancyhdr}       
\usepackage{graphicx}       
\graphicspath{{media/}}     

\title{Automated generation of large-scale distribution grid models based on open data and open source software using an optimization approach}

\author{ \href{https://orcid.org/0000-0002-1463-7606}{\includegraphics[scale=0.06]{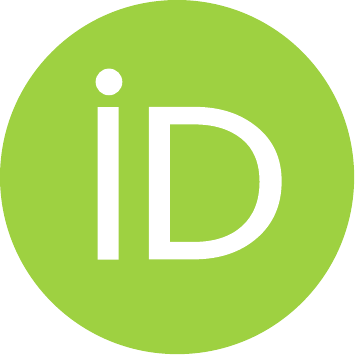}\hspace{1mm}H{\"u}seyin K. \c{C}akmak}\thanks{\url{www.iai.kit.edu}} \\
	Institute for Automation and Applied Informatics\\
	Karlsruhe Institute of Technology\\
	76344 Eggenstein-Leopoldshafen, Germany\\
	\texttt{hueseyin.cakmak@kit.edu} \\
	\And
	\href{https://orcid.org/0000-0002-1336-006X}{\includegraphics[scale=0.06]{orcid.pdf}\hspace{1mm}Luc Janecke} \\
	Institute for Automation and Applied Informatics\\
	Karlsruhe Institute of Technology\\
    76344 Eggenstein-Leopoldshafen, Germany\\
	\texttt{luc.janecke@student.kit.edu} \\
	\And
	\href{https://orcid.org/0000-0002-3572-9083}{\includegraphics[scale=0.06]{orcid.pdf}\hspace{1mm}Veit Hagenmeyer} \\
	Institute for Automation and Applied Informatics\\
	Karlsruhe Institute of Technology\\
    76344 Eggenstein-Leopoldshafen, Germany\\
	\texttt{veit.hagenmeyer@kit.edu} \\	
}

\begin{document}

\maketitle

\begin{abstract}
The increasing share of renewable energy sources on distribution grid level as well as the emerging active role of prosumers lead to both higher distribution grid utilization, and at the same time greater unpredictability of energy generation and consumption. This poses major problems for grid operators in view of, e.g., voltage stability and line (over)loading. Thus, detailed and comprehensive simulation models are essential for planning future distribution grid expansion in view of the expected strong electrification of society. In this context, the contribution of the present paper is a new, more refined method for automated creation of large-scale detailed distribution grid models based solely on publicly available GIS and statistical data. Utilizing the street layouts in Open Street Maps as potential cable routes, a graph representation is created and complemented by residential units that are extracted from the same data source. This graph structure is adjusted to match electrical low-voltage grid topology by solving a variation of the minimum cost flow linear optimization problem with provided data on secondary substations. In a final step, the generated grid representation is transferred to a DIgSILENT PowerFactory model with photovoltaic systems. The presented workflow uses open source software and is fully automated and scalable that allows the generation of ready-to-use distribution grid simulation models for given $20\,kV$ substation locations and additional data on residential unit properties for improved results. The performance of the developed method with respect to grid utilization is presented for a selected suburban residential area with power flow simulations for eight scenarios including current residential PV installation and a future scenario with full PV expansion. Furthermore, the suitability of the generated models for quasi-dynamic simulations is shown.
\end{abstract}


\textbf{\textit{CCS}}\enspace{\textbullet\enspace Computing methodologies → Model development and analysis; \textbullet\enspace Hardware → Renewable energy.}

\section{Introduction}
\label{sec: intro}
Amplified by political decisions worldwide that aim to reduce the emission of the greenhouse gas carbon dioxide and therefore mitigate climate change, a powerful trend towards electrification and distributed energy resources could be observed in recent years \cite{Zinaman.2018, Miller.2015}. The implications of this calls for a detailed analysis of the electrical grid on high-voltage (HV), medium-voltage (MV) and low-voltage (LV) level simultaneously that require a detailed computer model of the current grid. These models are also needed as a basis for a multi-modal energy system analysis that comprises electricity, gas and heating for a holistic investigation, including the interplay of distribution and transmission networks at the national level. However, unlike the power transmission system, distribution grid (DG) data is generally not freely available or is only partially available with restrictions.

This paper contributes by developing a methodology to tackle this lack of data for LV networks based on publicly available data that incorporate essential GIS information. Under the assumption that streets and public ways are possible cable routes, a graph structure of these streets and ways is extracted from OpenStreetMaps (OSM) \cite{OSMContrib.2004}, so that local geospatial conditions are taken into account. This graph structure is altered in a way that the resulting graph represents a radial electrical LV grid that can be transferred automatically to a DIgSILENT PowerFactory (PF) \cite{GonzalezLongatt.2014, DIgSILENT.2022} model.
In order to do so, a Python application is developed that uses an appropriate concatenation of the pre-existing, open source packages OSMnx, NetworkX, NumPy and Pyproj as well as the commercial optimizer Gurobi, that is free to use for academic purposes but can be replaced by any free suitable optimizer \cite{Boeing_OSMnx.2021, Hagberg.2008, NumPyDev.2021, Pyproj.2021, Gurobi.2021}. The developed application is tested in a suburban residential district with power flow and quasi-dynamic power flow simulations.

The paper is structured as follows. \Cref{sec:related_work} provides an  overview of related work. \Cref{sec:methodology} explains the methodology and \cref{sec:results} presents results that are discussed in \cref{sec:discussion} together with the proposed method. \Cref{sec:conclusion} concludes the paper.

\section{Related Work} \label{sec:related_work}
Using GIS based approaches to model an energy system has been proposed before. While \cite{Cakmak.2015} uses OSM to obtain a HV and MV grid model, \cite{Kays.2017} even adds the LV level and use OSM to extract building and street network data, which is stored in a graph structure for optimal secondary substation placement. Load estimation considers a building's footprint area without the building type and number of floors. With an assumed initial number of centroids as secondary substations and a distance matrix, a cluster analysis is performed. A fitness function coupled with load flow calculations is used to evaluate and optimize the network graph. The final grid topology is not radialized but the determined optimal position of the substation is evaluated with real data from 150 buildings and a few streets. In the context of modelling LV networks, a graph structure obtained from the street-layout was already proposed in \cite{MateoDomingo.2011}, where a reference network model with different topologies is created. Secondary substation locations need to be know in advance. In \cite{Navarro.2009}, optimal secondary substation locations are obtained together with a loss optimized grid but the approach focuses on planning new grids rather than depicting real ones. In \cite{Gouin.2015} a methodology is described to expand an existing distribution grid using the street-layout and a graph representation thereof. With the aim of extending the OSM and graph based research to other parts of the world, \cite{Ali.2020} develop a method to create grid dataset that can be used for benchmarking purposes but only considers the few secondary substations that are present in the OSM dataset. In the work of \cite{Sarajlic.2019}, that also uses the algorithm proposed by \cite{Kays.2017}, a methodology is presented to create benchmark models based on OSM data. Planning and operation principles are taken into account and with an iterative algorithm the LV network is adjusted to meet these principles. In \cite{Nasirifard.2018} a crowdsourcing approach for the collection of open data on DG devices with the aim of derivation of the grid topology is presented.
Our approach enables an automated generation of simulation ready distribution grid models and offers a more realistic assessment of distribution system operators (DSO) operating principles by taking into account their financial interests. The cost optimization does not require load flow calculations in order to meet these principles. Technical requirements and standards are considered in the proposed workflow, which is based on open source software and generates simulation models built on open data for commercial simulation software.

\section{Model Generation Methodology} \label{sec:methodology}
In the following, the methodology for generating a simulation ready low voltage grid model is explained. \Cref{sec:data_acquisition} deals with the automatic extraction of the data the model relies on. \Cref{sec:topology_extraction} focuses on converting the obtained structure to a grid topology, which is further processed with a network flow optimization to create radial topologies in \cref{sec:cost_optimization}. \Cref{sec:component_types} presents the parameter settings for the PowerFactory model to enable power flow and quasi-dynamic simulations.

\subsection{Data Acquisition}\label{sec:data_acquisition}
Apart from the street network of a given area to serve as potential cable routes, the grid model also needs data on buildings to model loads, batteries and PV generation. While all of that data is extracted automatically, the model still relies on given data about secondary substation locations and precise building characteristics. The chosen source for all GIS data is OpenStreetMap, the current PV deployment is obtained by analyzing satellite images from Bing Birdseye and potential PV capacity is obtained via governmental solar registers \cite{SolarkasterKA.2022}.
With OSMnx, a directed graph structure is created from the OSM raw dataset, that includes every way that OSM characterizes as publicly walkable. Building GIS data, such as its coordinates and its OSM tags is extracted from OSM using an html query to the Overpass API \cite{OverPass.2022}. Only buildings with an address and house number are considered. Further building data and secondary substation locations are gathered during an on-site inspection. 
All building data is collected in a database, where PV deployment and potential are mapped to the OSM building ids. From the raw data a buildings center coordinate, its footprint area, its yearly electricity consumption, its peak load share and the number of PV modules that can be constructed on each roof are calculated. 

\subsection{Grid Topology Extraction}\label{sec:topology_extraction}
In a further step the electric grid is assembled. Before each building can be augmented to the graph structure, the graph is converted to an undirected graph and the closest edge to each buildings center coordinate is calculated using a k-d tree for Euclidean nearest neighbor search. 
Each building is connected to the closest edge using orthogonal projection onto the edge.
Similarly, the secondary substations are added to the graph structure. To obtain the meshed grid structure, a simplification algorithm discards all intermediate nodes.
These are removed iteratively and replaced by a single edge that preserves the original paths geometry.

\subsection{Cost Optimization}\label{sec:cost_optimization}
In the next step, the meshed structure is converted to a radial structure using grid topological standards and rules, the most important of which is the permissible underground cable loading. As a distribution grid operator will always maximize profitability, the total length of all cable laid is minimized while at the same time supplying all demand and respecting cable capacity limits. This leads to a variation of the minimum cost network flow problem whose general form is extended by a radiality constraint. This guarantees that each demand is supplied by exactly one substation entirely. The optimization step requires the replacement of each undirected edge of the network by two identical edges with opposite directions.
To allow each secondary substation to supply a flexible amount of power, a virtual substation - the equivalent to an external grid element in PF - is inserted that supplies all secondary substations. These connections have zero length and are not considered in the minimization. This particular integer linear program takes a graph $G(V, E)$ as an input, where $V$ is the set of all graph nodes and $E$ the set of all edges. The model can then be stated using  \Cref{eq:formel1}-\ref{eq:formel4}.

\begin{equation}
\label{eq:formel1}
min \sum_{(i,j) \in E} c_{i,j} \cdot install_{i,j}
\end{equation}

\begin{equation}
\label{eq:formel2}
\sum_{j:(i,j) \in E} flow_{i,j} \quad - \sum_{j:(j,i)\in E} flow_{j,i} = s_i \quad \forall i \in V
\end{equation}

\begin{equation}
\label{eq:formel3}
\sum_{i:(i,j) \in E} install_{i,j}  \le 1 \quad \forall j \in V
\end{equation}

\begin{equation}
\label{eq:formel4}
0 \le flow_{i,j} \le s_{max,i,j} \cdot install_{i,j} \quad \forall (i,j) \in E
\end{equation}

The binary decision variable $install_{i,j}$ equals 1 if the corresponding edge from $node_i$ to $node_j$ is part of the solution, which means being used to supply buildings with power, and 0 in all other cases. The integer decision variable $flow_{i,j}$ represents the flow of power from $node_i$ to $node_j$. \Cref{eq:formel1} minimizes the total cost by adding up the costs $c_{i,j}$ of all cable sections that are used for power delivery. These costs are proportional to the line segments length. \Cref{eq:formel2} defines the flow conservation constraints that assure for each node, that all power that flows into it also needs to flow out or be consumed. Each node that represents a building has a negative consumption $s_i$, while the substation has a positive supply of power. All junction nodes where the cable routes split have a demand of 0 and therefore pass all power to the next line segments.
\Cref{eq:formel3} sets the radiality constraints that ensures for each node that at most one cable can be used to deliver power into the node. This translates to a radial structure as any cycle requires more than one input edge for at least one node. \Cref{eq:formel4} is defined to link the install and flow variables and express that no flow can exist over an edge that is not being built and even if it is built, it has to satisfy the maximum permissible cable loading $s_{max, i, j}$.
The solution of this model represents the cheapest way in which a DSO could supply all demand without exceeding the cable limits. In reality DSOs plan with a lot of reserve capacity so that the desired solution is the one with the lowest maximum cable loading, that is still able to satisfy all demand. 
The Newton-Bisection \cite{Mazhar.2016} method is applied to repeat the optimization while at the same time the permissible cable loading limit is lowered in order to find the lowest cable limit, with which the model remains feasible. 
It has to be mentioned that this optimization only considers cable loading from the consumption of buildings and does not consider potential PV modules or batteries installed. Furthermore, voltage characteristics, whose compliance is another important rule for LV grid topology are not considered. 

\subsection{Grid Component Types}\label{sec:component_types}
To transfer the radial grid structure to a PF model using its Python API, each calculation relevant parameter of each model component must be defined. See \Cref{app:component_types} for the used equipment types from the PF library.

\subsubsection{Cable Parameters}
As the model at this point only incorporates one cable strand per street while in reality, each side of the street would likely have its own, each cable that is not connecting to a building or a secondary substation is defined to be two parallel lines. While the building to grid connections remains a single strand of the standard cable, connections to secondary substations are set to be four parallel lines to consider the actual line ratings.
For the used standard types, the line lengths are set according to the underlying GIS model.  

\subsubsection{Transformer Parameters}
The used transformer models as given in \Cref{app:component_types} are used for the LV/MV and MV/HV grids. For the latter,  tap changers are defined for the HV side. This phase shifter adds 1,25\% of voltage per tap with a minimum position of -2 steps and a maximum of 2 steps.

\subsubsection{PV Panel Parameter}
As each roof either has several panels installed or the capacity to install more than one, these multiple panels are modeled with a single inverter. While in the present scenario each roof is equipped with the amount of panels that were built up until now, the future scenario assumes that 40\% of each roof is fitted with solar panels. In order to be able to have a time dependent model, each PV modules output is calculated by PF using the built-in solar calculation. Based on irradiance models, the time and place and therefore the suns angle of incidence as well as the PV panels location, orientation and tilt, PF can calculate the exact output of each module at any time.

\subsubsection{Load Parameter}
To accurately model the loads, each buildings households and their yearly consumptions are aggregated to a buildings yearly consumption $E_{b,a}$. The load estimation in \Cref{eq:formel5} is based on a household's number of residents $N_{R}$, the household's available area $A_{H}$ and the statistical number 8.4 of big electrical appliances per household for the selected test area. 
\begin{equation}
\label{eq:formel5}
E_{b,a} = N_{H} \cdot (N_{R} \cdot200\,kWh + A_{H} \cdot 9\,kWh + 8.4 \cdot 200\,kWh)
\end{equation}

For multi-storey buildings with more than one flat per story, each household of the total $N_{H}$ is assumed to have the same size. For purely non-residential buildings, consumption is based solely on the area and the type of non-residential use. For the test area it is distinguished between childcare buildings with a specific consumption of $22 \,kWh/m^2a$, schools with $20 \, kWh/m^2a$ and office like buildings with $45.41 \,kWh/m^2a$. Mixed use is aggregated accordingly. Given a yearly consumption for each building and the household's aggregated standard load profile \cite{Meier.1999}, PF can calculate the exact power draw of each connection at any point in time. Since this paper concentrates on residential DG, industrial loads are not considered.  

\section{Results} \label{sec:results}
This paper presents two results. Firstly, the application of the workflow to generate an LV residential grid model and secondly, its evaluation by testing the current and a future scenario, namely, how predicted PV deployment in the future will affect the grids.

\subsection{Automated Model Generation} \label{sec:results_model}
\begin{figure}
    \centering
    \includegraphics[height=10cm]{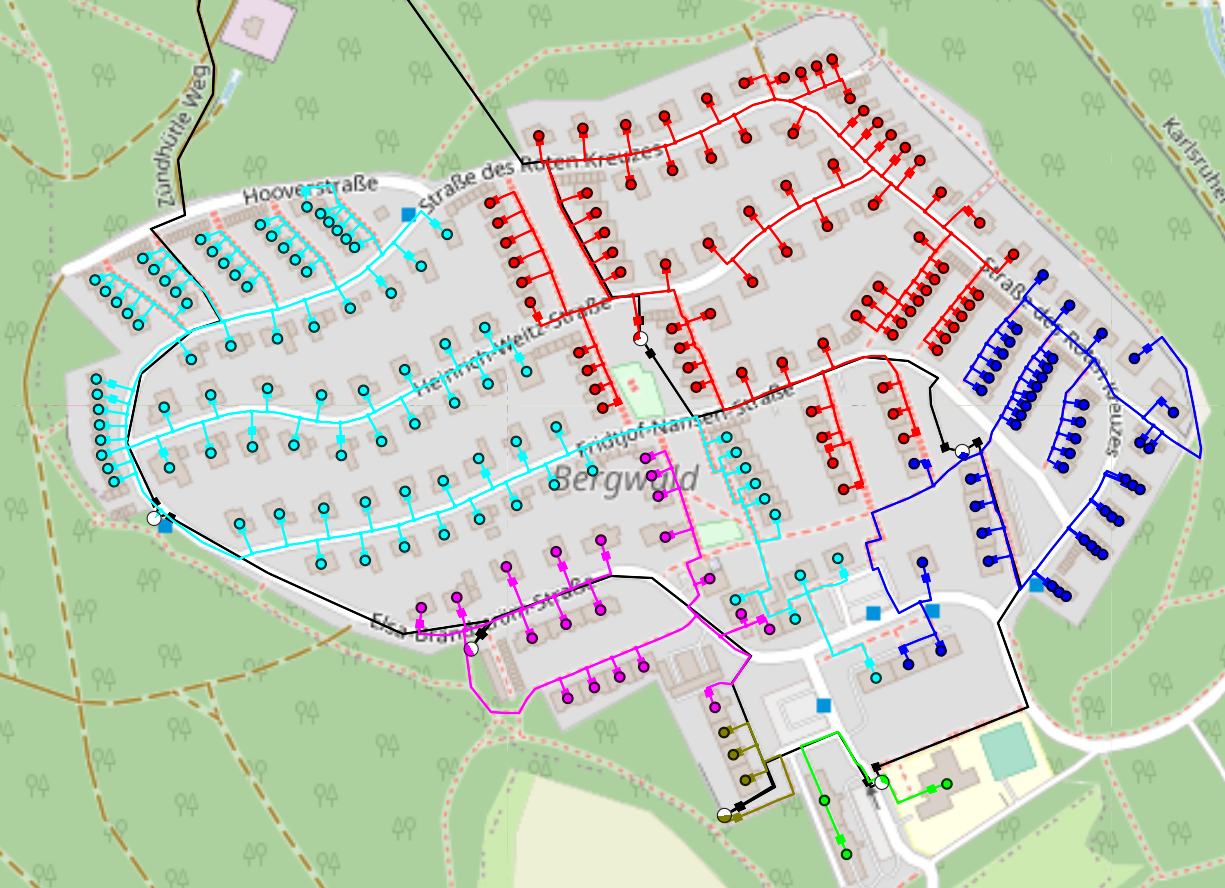}
    \caption{Topology of the distribution grid in PowerFactory as obtained by the introduced workflow.
    \label{fig:topo}}
\end{figure}

The workflow itself produces reasonable and reproducible results. In Figure \ref{fig:topo} the topology of an automatically generated distribution grid model in PF is shown.

This DG model consists of 241 site elements that represent houses, eight secondary substations, two of which do not supply the residential area but serve as connection bridges to the single substation that supplies the entire grid.
Furthermore there are 495 busbars, 3,435 terminals and 2,941 switches that account for the connection points of lines and busbar arrangements within buildings, secondary substations and substations. The grid model includes 485 lines, nine of which belong to the $20\,kV$ level, ten two-winding transformers, one for each secondary substation and two for the substation, 243 low-voltage loads, 241 of which represent each buildings load and one aggregated one for each of the two intermediary substations to also contain loads. Finally each house also incorporates a PV system, of whom only 10 are activated for the present PV scenario. Each distribution feeder is radial.  

\subsection{Model Evaluation} \label{sec:results_eval}
\begin{figure}
    \centering
    \includegraphics[height=10cm]{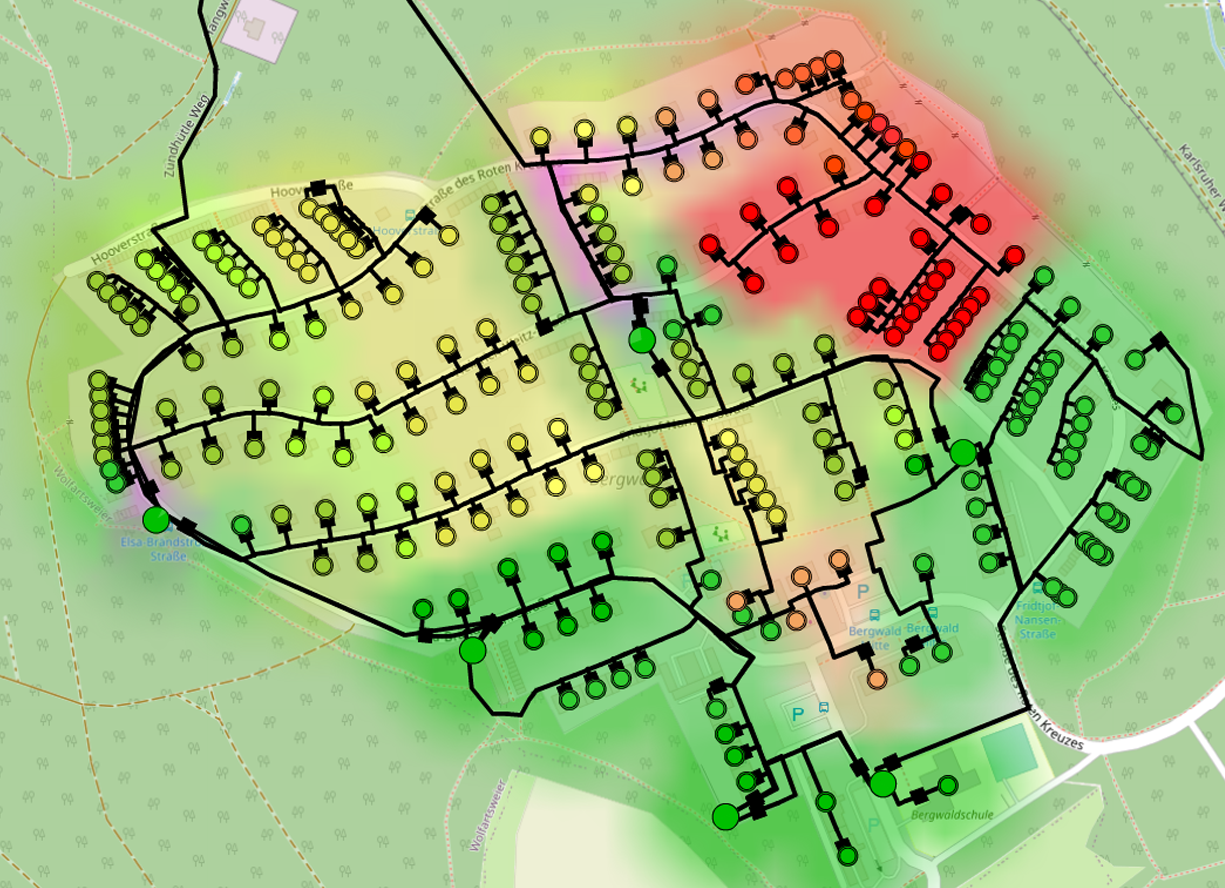}
    \caption{Power flow simulation results for line and transformer loadings and the bus voltages on July 15, 2021 12pm.
    \label{fig:simheatmap}}
\end{figure}

In order to evaluate the results of the workflow, 8 load flow calculations are carried out in PF. These load flows are selected to match a broad range of grid states and include the present PV deployment as well as the full scale PV deployment as an expected future scenario. For each of the two PV scenarios, a load flow is carried out for a sunny summer day with maximum radiation, the same summer day but during peak load and the same for a winter day. For this the July 15, 2021 and the December 15, 2021 were selected, both at 12pm and 8pm. The result of the load flow simulation in July 15, 2021 12pm is shown in \Cref{fig:simheatmap}.

Simulation results of all scenarios are presented in \Cref{app:sim_results_pf}. It becomes evident that in the present PV scenarios, at all times the grid is in a good state with line loading not exceeding 16.55\% and voltage not dropping below $0.977 p.u.$ The transformer loading of at most 18.41\% as well as the total losses of at most $14.94\,kV\!\!A$ are representing a stable distribution grid. Since the most extreme values are always in winter at about 20pm, which coincides with the scenario with the highest load, these results are consistent and realistic. 

Looking at the scenario with PV deployment up to its full capacity, another picture is revealed. While for the evening hours a similar behavior as in the previous case can be seen, during daytime the grid is highly overloaded. During winter, at 12pm this effect due to the lower temperatures is not yet catastrophic since line loading peaks at 66.23\% and the voltage does not exceed 1.051. In summer on the other hand equipment failures are inevitable with continuous line loadings of more than 120\%, voltage rises that with more than $1.1 p.u.$ violate all voltage characteristic standards and continuous transformer loadings of over 100\%. Additionally, $220.55\,kV\!\!A$ are lost in transmission. From this, one can clearly conclude that the present grid, even though currently in a good state, is not able to host all this PV injection without massive reinforcements or the large-scale implementation of storage systems with smart, adaptive control. 

As part of the verification process, a quasi-dynamic simulation is carried out for the weeks July 12-18, 2021 and December 12-18, 2021 with a time interval of 15 minutes for both, the present and future PV scenario. The results are given in \Cref{app:sim_results_qsd}. It becomes evident that the results are similar to the ones of the single power flows as they were selected in order to match the most extreme conditions. The overall maximum loading is about 141.47\% when the suns radiation is at its climax during summer at midday. The highest average line loading is 51.44\%, which indicates that the cable's loading is well above its limit for a long time given that during night the line's loading is very low.  

\section{Discussion}\label{sec:discussion}
The workflow introduced in the present paper bears the significant advantage of being applicable anywhere on earth with good OSM data coverage. Nonetheless, it still needs the exact secondary substation and substation locations as well as the $20\,kV$ line paths. However, due to the selection of a network flow problem for topology extraction, from an operations research point of view it is straightforward to also include the $20\,kV$ MV level into the optimization and model the secondary substation locations as a location problem. A heuristic estimation of the secondary substations is indispensable. For load estimation precision, more data sources than OSM can be considered, such as local cadastre and statistical data. It is also important to conclude that the number of storeys of a residential building are significant to get a good estimate of the number of households for load modelling.

The limitations of the proposed method mainly originate from the lack of real-world data and the assumed simplifications of the model, which do not consider switches and cable distribution cabinets at the current stage of development. Only one strand of cable per street, the same equipment type for any element of the same kind and non-residential use of buildings are only considered to a very small extend. Lastly, load modelling is simplified and assumed to be proportional to the number of residents, area, and number of large electrical appliances in a household. The optimization itself does not yet offer the functionality to also decide on bigger cables for a highly loaded route, so that many times a DSO would just reinforce their cables, the optimization instead finds a new, longer path to connect buildings to a secondary substation. 

\section{Conclusion}\label{sec:conclusion}
This paper introduces a new method for automated generation of simulation models of LV grids for suburban residential areas. OSM and publicly available data are used with open software to create simulation models that are suitable for power flow and quasi-dynamic simulations. For the evaluation of the developed methods, various weather conditions are simulated for the current and a future scenario with full PV expansion. The simulation results indicate heavy congestion on distribution grid level for full PV capacity deployment without batteries. 

Future work will focus on the removal of the aforementioned limitations and on the enhancement of the optimization methods for including secondary substation locations, the model integration of batteries with smart control strategies and scaling of the application in a high performance computing environment with the goal of co-simulation of energy systems including transmission grids and detailed distribution grids at national level.

\section*{Acknowledgments}
This work was conducted within the framework of the Helmholtz Program Energy System Design (ESD).

\bibliographystyle{unsrt}  
\bibliography{references}  

\appendix

\section{Appendix}

\subsection{PowerFactory Model of a Residential Unit}\label{app:residential_model}
Each of the 241 sites is modelled as a residential unit (house) that contains an aggregated load of the apartments and photovoltaics.
\begin{figure}[H]
    \centering
    \includegraphics[height=6cm]{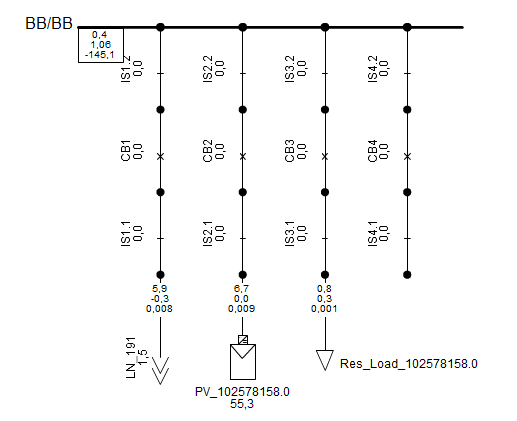}
    \caption{PowerFactory model of a residential unit comprising aggregated load of the apartments and photovoltaics.}
\end{figure}

\subsection{Model Components in PowerFactory}\label{app:component_types}
\begin{table}[H]
    \centering
    \caption{Used components from the PF equipment library.}
    \label{tab:QuasiPres}
    \begin{tabular}{*2{c}}  
        \toprule
        Model Type& Description  \\
        \midrule
        NAYY4x150SE 0.6/1 kV LV     & Cable for 0.4 kV LV grid\\
        NA2XS2Y1x95RM 12/20 kV      & Cable for 20 kV MV grid\\
        DTTHL SGB 0.63MVA 20/0.4 kV & Transformer for MV/LV\\
        DTTHL SGB 40MVA 110/20 kV   & Transformer for MV/HV\\
        Aleo S19.230                & PV 230W, area $1,6434m^2$\\
        \bottomrule
    \end{tabular}
\end{table}

\subsection{Power Flow Simulation Results in PF}\label{app:sim_results_pf}
\begin{table}[H]
    \centering
    \caption{Results of the power flow simulation for the current PV-installation in 2021.}
    \label{tab:fullPVSim}
    \begin{tabular}{*5{c}}  
        \toprule
        Measurement& Jul 12pm & Jul 20pm & Dec 12pm & Dec 20pm\\
        \midrule
        \normalsize{Trafo Loading max.}  & 12.74\%  & 16.52\%  & 10.86\%  & 18.41\%\\
        \normalsize{Line Loading max.}   & 11.22\%  & 14.80\%  & 9.76\%   & 16.55\%\\
        \normalsize{Line Loading avg.}   & 2.19\%   & 2.80\%   & 1.88\%   & 3.09\%\\
        \normalsize{Voltage max.}        & 1.00p.u. & 1.00p.u. & 1.00p.u. & 1.00p.u.\\
        \normalsize{Voltage min.}        & 0.98p.u. & 0.98p.u. & 0.98p.u. & 0.98p.u.\\
        \normalsize{Total Losses}        & 12.27kVA & 13.98kVA & 11.60kVA & 14.94kVA\\
        \bottomrule
    \end{tabular}
\end{table}

\begin{table}[H]
    \centering
    \caption{Results of the power flow simulation for a future scenario with full capacity rooftop PV deployment.}
    \label{tab:results_full_PF}
    \begin{tabular}{*5{c}}  
        \toprule
        Measurement& Jul 12pm & Jul 20pm & Dec 12pm & Dec 20pm\\
        \midrule
        \normalsize{Trafo Loading max.}  & 109.63\%    & 9.28\%   & 53.34\%  & 18.41\%\\
        \normalsize{Line Loading max.}   & 133.07\%    & 9.92\%   & 66.23\%  & 16.55\%\\
        \normalsize{Line Loading avg.}   & 16.35\%     & 1.50\%   & 7.83\%   & 3.09\%\\
        \normalsize{Voltage max.}        & 1.10p.u.    & 1.00p.u. & 1.05p.u. & 1.00p.u.\\
        \normalsize{Voltage min.}        & 0.99p.u.    & 0.99p.u. & 0.98p.u. & 0.98p.u.\\
        \normalsize{Total Losses}        & 220.55kVA   & 11.26kVA & 42.24kVA & 14.94kVA\\
        \bottomrule
    \end{tabular}
\end{table}

\subsection{Quasi-Dynamic Simulation Results in PF}\label{app:sim_results_qsd}
\begin{table}[H]
    \centering
    \caption{Results of the quasi-dynamic simulation for a summer and winter week for the present PV deployment.}
    \label{tab:results_current_quasi}
    \begin{tabular}{*5{c}}  
        \toprule
        Measurement&  Jul 12-18 &  Dec 13-19 \\
        \midrule
        \normalsize{Trafo Loading max.}      & 17,6\%    & 21,08\% \\
        \normalsize{Line Loading max.}       & 15,77\%   & 18,97\% \\
        \normalsize{Line Loading avg. max.}  & 8,95\%    & 9,15\%  \\
        \normalsize{Voltage max.}            & 1.00p.u.  & 1.00p.u.\\
        \normalsize{Voltage min.}            & 0.98p.u.  & 0.97p.u.\\
        \bottomrule
    \end{tabular}
\end{table}

\begin{table}[H]
    \centering
    \caption{Results of the quasi-dynamic simulation for a summer and winter week for a potential full capacity rooftop PV deployment.}
    \label{tab:results_full_quasi}
    \begin{tabular}{*5{c}}  
        \toprule
        Measurement& Jul 12-18 &  Dec 13-19 \\
        \midrule
        \normalsize{Trafo Loading max.}      & 116,99\%  & 53,57\%  \\
        \normalsize{Line Loading max.}       & 141,47\%  & 66,43\% \\
        \normalsize{Line Loading avg. max.}  & 51,44\%   & 19,47\%  \\
        \normalsize{Voltage max.}            & 1.118p.u. & 1.05p.u. \\
        \normalsize{Voltage min.}            & 0.98p.u.  & 0.97p.u. \\
        \bottomrule
    \end{tabular}
\end{table}

\end{document}